\documentclass[preprint]{aastex}

\newcommand{\bq}{\begin{equation}}
\newcommand{\eq}{\end{equation}}
\newcommand{\h}{$^h$}
\newcommand{\m}{$^m$}
\newcommand{\s}{$^s$}
\newcommand{\dg}{$^\circ$}
\newcommand{\p}{$\pm$~}
\newcommand{\am}{$'$}
\newcommand{\as}{$''$}

\begin{document}
%\vskip 0.6 in
%\noindent
 
\title{Proper Motion Measurements with the VLA: I. Wide-field Imaging and Pulse Gating Techniques}
 
\author{
R. S. McGary\altaffilmark{1},
W. F. Brisken\altaffilmark{2},
A. S. Fruchter\altaffilmark{3},
W. M. Goss\altaffilmark{4} and 
S. E. Thorsett\altaffilmark{5}}

\altaffiltext{1}{Harvard-Smithsonian Center for Astrophysics, Mail Stop 10, 60 Garden Street, Cambridge, MA 02138}
\altaffiltext{2}{Department of Physics, Princeton University, Princeton, NJ 08544}
\altaffiltext{3}{Space Telescope Science Institute, 3700 San Martin
Drive, Baltimore, MD 21218}
\altaffiltext{4}{National Radio Astronomy Observatory, PO Box O, Socorro, NM 87801}
\altaffiltext{5}{Department of Astronomy and Astrophysics, University
of California, Santa Cruz, CA 95064}

\slugcomment{Accepted by Astronomical Journal on 18 October, 2000}

\begin{abstract}
The pulsar velocity distribution provides information about the binary
  history of pulsar progenitors as well as asymmetries of the supernova
  events in which pulsars are born.  Studies of local pulsars present a 
  biased view
  of this distribution, since they preferentially select low velocity
  pulsars that have remained near their birthplaces in the Galactic plane.
  Using the VLA, we have studied the proper motions of a large
  sample of distant pulsars.  These pulsars are generally faint,
  and the expected proper motions are small.  In this paper, we describe
  the data analysis techniques that we have developed to allow
  precise astrometric measurements of faint sources with the VLA.  These
  techniques include ``gating'' the VLA correlator to increase the
  signal-to-noise ratio of the pulsar by gathering data only during the
  pulse.  Wide-field imaging techniques, including multiband imaging to
  account for bandwidth smearing, were used to allow identification of
  multiple in-beam background sources for astrometric calibration.  We
  present the analysis of three pulsars, and demonstrate that astrometric 
  accuracy of about ten milliarcseconds can be obtained for individual sources with our technique, allowing measurement of proper motions with errors of only a few milliarcseconds per year over our seven year baseline.

\end{abstract}

%\vfill
%\eject


\section{Introduction}

Interferometric proper motions have always been hindered by the low
flux densities of many pulsars.  This restriction results in a bias in
which only the closest  pulsars are observed.  Although these pulsars move quickly on angular scales, their absolute motion must be small for them to still be near the plane of the Galaxy.
The fastest pulsars, however, have traveled the greatest
distances and may be more than a kpc above the Galactic plane.
Previous interferometric proper motion projects have been unable to
include these distant pulsars due to their low flux densities.  The
study by Fomalont et al., begun in 1984, was limited to pulsars with
$S>2$ mJy so that an adequate signal to noise could be achieved
\citep{fom92}.  Few faint pulsars have proper motion measurements as
a result of these selection effects.

This study of proper motions of distant, high $z$ pulsars using the
Very Large Array (VLA) of the National Radio Astronomy
Observatory\footnote{The National Radio Astronomy Observatory is a
facility of the National Science Foundation operated under cooperative
agreement by Associated Universities, Inc.} began in 1992.  We have
chosen a group of 28 pulsars, 15 of which have $|z|>$ 400 pc, $d\le
4000$ pc, and $|b| < 30^\circ$.  This selection criteria minimizes the 
uncertainty in the determination of the proper motion in the $z$ 
direction which enables the study of proper motions of pulsars
relative to the Galactic plane.  Upon
completion, this project will double the number of proper motions
measured for distant, high $z$ pulsars.  The flux density restriction
is overcome by gating the VLA.  In order to have the pulsar in the
same field as stationary sources, we must create wide-field images
with position accuracies of a few mas.  This requirement forces us to
account for small effects such as annual aberration and Lorentz
contraction over the fields.

This paper focuses on the method of data reduction and error analysis
used to calculate the proper motions.  Three pulsars (B0919+06,
B1237+25 and B1937--26) are used as ``case studies'' demonstrating the
accuracy of the method over a range of declinations and flux densities.
Proper motions for all 28 pulsars will be presented by Brisken et al.
 in Paper II.

\section{Data and Observational Techniques}

Five epochs of data centered on 1992.96, 1994.23, 1995.52, 1998.32 and
1999.47 have been gathered at the VLA, amounting to more than 120
hours of observations.  Every pulsar has been observed twice; 25
pulsars have three or more epochs of data.

\subsection{{\it Observational Setup}}

Observations were made in the A array at the VLA with the correlator
in the 2 AD mode.  A maximum baseline of 36 km provides the highest
possible resolution (1.1$''$).  With this resolution, large sources
are resolved out leaving mostly distant, extra-galactic sources in the
images.  These sources should be stationary over time and are used as
reference sources against which to measure the motion of the pulsar.

The data were gathered at 20 cm (1452.4 MHz) as a compromise between the reduction
in pulsar flux density at higher frequencies and decreased resolution
at lower frequencies.  A 25 MHz bandwidth was divided into fifteen
channels ($\Delta\nu=1.56$ MHz) to provide the large usable field of
view ($\approx36'$) necessary for wide-field imaging (see section
2.2).  Observations were made in two circular polarizations to provide
the gating capability for pulsars without losing the ability to image
weak continuum sources (see section 2.4).  Thirteen channels were used
to produce final images (see section 3).

Each pulsar was typically observed four times for 15 minutes during an
observing run.  Each segment was bracketed by five minutes of
observation of a bright nearby VLA calibrator that was subsequently
used for phase calibration and secondary amplitude calibration.

\subsection{{\it Delay Beam $\&$ Bandwidth Smearing}} The size of the
primary beam of the VLA is given by $\theta_p\approx\lambda/d$ where
$d=25$ m is the diameter of the dish.  Thus $\theta_p\approx30'$ at 20
cm defines the field of view contained inside the first null of the
primary beam.  In practice, observations are made over a band of
non-zero width ($\Delta \nu \ne 0$). The range of frequencies in the
band results in bandwidth smearing in which sources far from field
center are smeared in the radial direction.

Bandwidth smearing is characterized by the delay beam, which is
defined as $\theta_d=2c/\Delta\nu D$ where $D$ is the maximum baseline
(36 km for the A array).  To maximize the usable field of view,
observations are set up so that $\theta_d\approx\theta_p$.  Since a
single continuum channel ($\Delta \nu =25$ MHz) has a usable field of
view of $\sim$2$'$, observations must be made in the spectral line
mode with narrow channels.  By dividing the band into fifteen channels
($\Delta \nu =1.56$ MHz) a usable field of view of $\sim$36$'$ is
attained.  Bandwidth smearing is only an important effect for sources
more than $\sim$18$'$ from field center.  The added uncertainty in
position of reference sources more than 20$'$ from field center is
taken into account in the proper motion fitting program (see section
6)

\subsection{{\it Gating}}

The VLA correlator was gated using the Princeton Mark III Pulsar
Timing Machine \citep{sti92}.  This computer allows real-time
adjustment of the pulsar gate enabling the correlator to record data
only when the pulsar is ``on.''  All the data during the off-pulse are
discarded and only on-pulse portions are retained.

Gating can increase the signal-to-noise ratio (SNR) of a pulsar by up
to a factor of five.  The gate is matched to the duty cycle of the
pulsar to maximize the improvement in the SNR.  This improvement is
approximately proportional to the inverse square-root of the duty
cycle.  Because the VLA has only a single gating circuit for all
channels, the width of the gate must account for dispersion smearing
across the full 25 MHz bandwidth and typically ranges between 5 and
15\% of the period for pulsars in this study.  For B0919+06, gating
increased the flux density from 6 to 120 mJy while the noise went from
0.1 to 1.5 mJy.  B1237+25 showed an increase in flux density of 5 to
55 mJy and an increase in noise of 0.2 to 0.7 mJy.  Finally, the flux
density of B1937--26 increased from 1 to 15 mJy while the noise
increased from 0.2 to 0.6 mJy when the VLA correlator was gated.
Therefore, gating increased the SNR of B0919+06 by a factor of
$\sim$1.3 while B1235+25 and B1937--26 both showed an increase in SNR
of a factor of $\sim$4.  In general, gating is not applied to pulsars
that are brighter than typical reference sources ($\sim10$ mJy).
However, technique testing (such as LL, RR alignment) was done using
gated, bright pulsars while hardware availability prevented gating on
some weak pulsars.

For our observations, the VLA is gated in the right circular
polarization (RR) in order to increase the SNR of the pulsar.  The
left circular polarization (LL) is not gated, to permit optimal
detection of reference sources in the field.  Since the position of
the pulsar is compared to those of the reference sources, positions in
the right and left polarizations must agree to a few mas.  Positions
of strong pulsars and reference sources show a typical agreement
between polarizations of 3 to 5 mas.

\section{Calibration}

The processing of the data begins with the removal of discrepant data
points using the procedure TVFLG in $AIPS$ (Astronomical Image
Processing System).  During the first observation of a pulsar, the
first 3 to 5 minutes are typically used to determine the optimum
position of the gate.  Since the ungated data have significantly lower
SNR than the gated data, they must be removed from the data set.

Calibrators for each pulsar are positioned at the phase center and are
unaffected by bandwidth smearing.  Therefore, basic calibrations are
done in the continuum data set, which is a sum of the inner 75\% of the observing band, and the resulting tables are copied to
the spectral line data.  3C286 (J1331+305), 3C48 (J0137+331) and 3C147
(J0542+498) were used as the flux density and bandpass calibrators for
the observations.  Channels on the edge of the band pass filter are
degraded by a roll-off in the band which results in reduced sensitivity
in these channels.  The first and last channels were removed from the
data leaving thirteen channels to be used for final images.
Calibration and editing information was applied when splitting the
{\it uv} data into separate data sets for each pulsar.

For the clean algorithm to converge properly, the brightness and
structure of all sources must remain constant, in both time and
frequency, over the entire integration.  The regular amplitude
calibration and removal of data taken without the gate ensure that the
time constancy is achieved, and the bandpass calibration flattens the
spectrum.  Although pulsars have considerably steeper spectra than do
most of the reference sources and calibrators, their mean flux density
changes by less than 5\% over our band and produces no noticeable
cleaning artifacts.  Scintillation in frequency and time
violate the source constancy that is assumed in the clean algorithm,
resulting in cleaning artifacts and an increase in the background noise.
Since the noise near each source is used to determine
the uncertainty of its position fit, this source of uncertainty is
automatically included in the analysis.

\subsection{{\it UVFIX}}

Before imaging, UVFIX is run on the single source data sets.  This
$AIPS$ program recovers the correct $u, v$ and $w$ coordinates of a
source that are only approximated by the VLA correlator.  The omission
of these terms results in a movement in the tangential direction of up
to 60 mas for a source 10$'$ from the field center \citep{fom92}.
UVFIX also corrects for Lorentz contraction of the field due to the
motion of the Earth along the line of sight. The magnitude of this
effect depends on the time of year in which the observations were made
and can result in a maximum radial movement of 60 mas for a source
10$'$ from field center \citep{fom92}.

L. Kogan at NRAO-Socorro revised UVFIX to account for these two
effects by recalculating the correct $u, v$ and $w$ coordinates as
well as moving to a stationary frame relative to the Sun. The results
of the updated UVFIX program were tested by comparing reference source
position agreement between epochs before and after UVFIX was run.  As
seen in Figure \ref{uvfix:plot}, UVFIX removed the rotation in the
field near B0919+06 improving reference source alignment by a factor
of three.  This new version of UVFIX was incorporated into {\it AIPS}
in early 1998.

\section{{ Imaging Techniques}}

Images are made using the Clark
``clean'' algorithm in $AIPS$.  A pixel size of $0.15''$ provides more
than 5 pixels/beam.  Clean boxes are placed around regions containing
sources to ensure that sidelobes are removed from these areas.
Separate clean boxes are used for the pulsar and each reference
source.  A robust weighting of zero, which indicates a compromise
between natural and uniform weighting, is used for all imaging.

\subsection{{\it Wide-field Imaging}}

A wide-field image ($60'\times 60'$) is made in the LL polarization
with ungated data for the detection of reference source candidates.
Positions of bright sources are recorded and new images are made with
separate fields for each source.  Ideally, only point sources are
retained as reference sources.  In cases where there are only a few
point sources we choose slightly extended sources as additional
reference sources.  The larger error in the position estimates of
these extended sources results in additional uncertainty in the final
proper motion.  All sources, including extended and weak sources are
included in the imaging process to remove sidelobes from other fields.
Due to the wide field used to image the reference sources and the
non-coplanarity of the VLA, the {\it uv} coverage at different points
within the field of view can differ by an appreciable amount \citep{per99}.  
To account for this, the pulsar and all reference sources are
imaged using {\it uv} values appropriate for their part of the sky.

Multiple reference sources have been found for all 28 pulsars.  There
are typically about eight good reference sources for each pulsar.  The
quality of each reference source is determined by measuring its proper
motion relative to the other reference sources.  Extended reference
sources are generally not used and point-like reference sources with
large motions are also omitted.  Sources used as reference sources for
the final proper motion calculations have their proper motions listed
in Table \ref{refsrc}.  For the three pulsars considered in this
paper, the maximum distance of any detected source from field center
is 27.6$'$.  However, no source further than 22.3$'$ was used for the
proper motion calculations.

\subsection{{\it Self-calibration}}

If there is enough flux density in compact sources inside the inner
third of the primary beam then self-calibration can improve the
signal-to-noise ratio of the detections.  In self-calibration, one
assumes that the image is degraded by antenna based gain and phase
errors which vary in time and prevent perfect calibration.  By
re-calibrating the data using an initial set of images as a model,
corrections to these phase and amplitude errors can be calculated as a
function of time.

An accurate model of the flux density distribution across the field is
necessary for self-calibration to be successful.  Self-calibration is
performed in the left hand polarization since the left hand data
typically contain more flux density than the gated right hand
polarization.  In addition, pulsar scintillation is much more
prominent in the gated polarization and could introduce additional
amplitude errors in the calibration.  Solutions from the left hand
data are applied to both polarizations.  Self-calibration can result
in a shift in the field of as much as 20 mas when the initial model is
not complete.  Since the pulsar and reference sources are equally
affected by this shift, the proper motion calculations are not
compromised.  Self calibration and imaging loops were iterated between
2 and 4 times.  Solution intervals of between two and five minutes
were used.

\section{Position Determination}

Positions of the pulsar and reference sources are found from final
images using the {\it AIPS} task JMFIT.  This task fits a gaussian
profile to a point source with a width based on the size of the
synthesized beam.  For extended sources, the program solves for the
width of the source as well.

JMFIT also reports an uncertainty in the position estimate.  Since
this uncertainty directly affects our confidence in the proper
motions, tests were run on the output of JMFIT to ensure that it gives
a reasonable error estimate.  A continuum point source in 41 line-free 
channels of an HI image
of NGC4688 was used for this test.  These data were chosen because a
bright point source and a large number of channels were available.
Using identical parameters, JMFIT was run on each channel individually
and the position and error output were recorded.  The uncertainties
reported by JMFIT agreed with the standard deviation of the 41
positions reported for the point source.

\section{Proper Motions} \label{sec:motions}

The pulsar positions and proper motions are determined through a
global least squares fit.  Positions of reference sources and shifts
in the coordinate system between epochs are also fit.  The beam shape
is used with the uncertainty reported by JMFIT to produce a correctly
oriented elliptical gaussian uncertainty.  An additional uncertainty
of $F \cdot R \cdot \Delta\nu/\nu$ is added in quadrature in the
radial direction to account for bandwidth smearing where $R$ is the
source's distance from the phase center, $\Delta\nu/\nu$ is the single
channel fractional bandwidth, which in our case is close to 0.001, and
$F$ is an empirically determined constant. By measuring the scatter in
position measurements of point-like reference sources far from the
field center in the same way that JMFIT was tested, we have determined
that $F=0.08$ for our data.

A small systematic coordinate offset remains for each epoch, even
after UVFIX is applied.  This effect ($\le$30 mas for a source 10$'$
from field center) is approximately half as large as the correction
made by UVFIX and cannot be adequately modeled by a simple rotation
and dilation.  Although the source of this offset is not fully understood, 
it is effectively removed by fitting for a six
parameter general linear transformation, $\vec{X}^{\prime} =
{\mathbf{A}} \vec{X} + \vec{B}$, between the coordinate systems of
each epoch and the first epoch.  $\mathbf{A}$ is a two by two matrix
whose elements typically deviate from the identity matrix by a few
times $10^{-5}$ and $\vec{B}$ is a coordinate frame shift between the
two epochs.  The amount of correction can be
characterized by a dimensionless number equal to the RMS value of the
matrix elements.  If we define $N_{epochs}$ as the number of epochs in which 
the pulsar was observed and $\delta {\mathbf{A}}_{ij} =
{\mathbf{A}}_{ij} - \delta_{ij}$, where $\delta_{ij}$ is the identity
matrix, then

\begin{equation}
\mathrm{RMS} = \sqrt{\frac{1}{4*(N_{epochs}-1)} \sum_{e=2}^{N_{epochs}} 
(\delta {\mathbf{A}}_{e;11}^2 +
\delta {\mathbf{A}}_{e;12}^2 +
\delta {\mathbf{A}}_{e;21}^2 +
\delta {\mathbf{A}}_{e;22}^2)
}\phm{000}.
\end{equation}

The RMS correction for 28 pulsars is plotted against source
declination in Figure \ref{fig:elevation}.  Since most of the
observations were made near transit the elevation is approximately $90
- \left|dec - lat \right|$, where $lat$ is the lattitude of the VLA,
about $+34^{\circ}$.  The required correction seems to follow the
secant of the elevation suggesting that the atmosphere may cause this
effect.

A Monte-Carlo bootstrap method was used to better determine the final
error ellipses.  In this test, data are randomly resubstituted and the
solution is fit thousands of times.  The solutions for $\mu_{\alpha}$
and $\mu_{\delta}$ are plotted in a scatter plot.  The uncertainties are
estimated from the width of the distribution with an orientation based
on the shape of the distribution in right ascension and declination.
The uncertainties obtained in this manner agree well with the least
squares errors.  For a complete description of this method, see {\em
Numerical Recipes in C} \citep{pre92}.

\section{Case Studies}

B0919+06, B1237+25 and B1937--26 demonstrate the proper motion
calculation as well as the accuracy of the error analysis.  B0919+06
and B1237+25 were chosen because they both have previously published
proper motions and can be used to test our results against those
obtained from other methods.  B0919+06 is especially interesting
because it has a VLBA determination of the proper motion accurate to
$<1$ mas yr$^{-1}$ \citep{chapre}.  B1937--26 is included because it
is a weak pulsar ($\sim$1 mJy) and is also located at a low
declination.  At --26$^\circ$, the synthesized beam of the VLA is
2.5$''$ in declination compared to only 1.1$''$ in right ascension.
This elongated beam makes accurate measurement of $\mu_\delta$
difficult.  In addition, calibrations for low elevation sources depend
strongly on pointing direction.  Depending on the distance from pulsar
to phase calibrator, this can make self-calibration difficult.  The
calculated proper motions for all three pulsars can be seen in Table
\ref{pm:table}.

\subsection{{\it B0919+06}}

B0919+06 provides the most stringent test of the accuracy of our
method.  We initially compared our measurement of the proper motion
($\mu_\alpha=$18.8 \p 0.9 mas yr$^{-1}$, $\mu_\delta=$86.4 \p 0.7 mas
yr$^{-1}$) to that published by \citet{fom99} ($\mu_\alpha=$17.7 \p
0.3 mas yr$^{-1}$, $\mu_\delta=$79.2 \p 0.5 mas yr$^{-1}$).  There was
an obvious disagreement between the two results, especially in
$\mu_\delta$.  A recent analysis of more extensive VLBA data (Fomalont
et al. data plus additional VLBA observations from October,
1998) by Chatterjee, Fomalont et al. (in preparation) has led
to a revision of the initial VLBA result.  The revised VLBA values
are: $\mu_\alpha=$18.4 \p 0.2 mas yr$^{-1}$, $\mu_\delta=$86.7 \p 0.3
mas yr$^{-1}$ with a parallax of $\pi=$1.15 \p 0.25 mas
\citep{chapre}.  We agree with this result to less than 1$\sigma$
in $\mu_\alpha$ and $\mu_\delta$.  It is important to note that the
errors from our new technique are comparable to those of the updated
VLBA observations.  The new measurement also agrees to within
1$\sigma$ of the proper motion measurement by \citet{har93} 
of $\mu_\alpha=$13 \p 29 mas yr$^{-1}$,
$\mu_\delta=$64 \p 37 mas yr$^{-1}$.  Figure \ref{pm:plot}a shows the
new proper motion measurement along with the \citet{chapre} and 
\citet{har93} results.

\subsection{{\it B1237+25}}

The proper motion measured for B1237+25 ($\mu_\alpha=$--104.5 \p 1.1
mas yr$^{-1}$, $\mu_\delta=$49.4 \p 1.4 mas yr$^{-1}$) also agrees
with both previous measurements (see Figure \ref{pm:plot}b).  Our
measurement deviates by $<$1$\sigma$ from the \citet{fom92} result 
of $\mu_\alpha=$--113 \p 13 mas yr$^{-1}$,
$\mu_\delta=$43 \p 14 mas yr$^{-1}$.  It also agrees reasonably well
with the \citet{lyn82} measurement of
$\mu_\alpha=$--106 \p 4 mas yr$^{-1}$, $\mu_\delta=$42 \p 3 mas
yr$^{-1}$.

\subsection{{\it B1937--26}}

A proper motion of $\mu_\alpha=$12.1 \p 2.4 mas yr$^{-1}$,
$\mu_\delta=$--9.9 \p 3.8 mas yr$^{-1}$ was measured for B1937--26
(see Figure \ref{pm:plot}c).  Due to the low flux density of this
pulsar ($\sim$1 mJy), B1937--26 has no previous measurement of its
proper motion.  Despite the elongated beam at low declination, the
value for $\mu_\delta$ has a small error.  The motion of the pulsar is
significant when compared to the small motion of its reference sources
(see Table \ref{refsrc}).

Without gating, we would not be able to determine the proper motion of
this pulsar to such high accuracy.  The proper motion of B1937-26 was
recalculated using only the ungated, LL polarization for the positions
of the pulsar and the reference sources.  The resulting proper motion
was $\mu_\alpha=$20 \p 8 mas yr$^{-1}$, $\mu_\delta=$--15 \p 15 mas
yr$^{-1}$.  The increase in error due to the low signal to noise of
the pulsar makes the measurement much less significant than the gated
case.

The successful proper motion measurements for these three pulsars show
the accuracy of this technique.  B1937-26 is also limited by low
flux density.  A precise proper motion measurement would have been
impossible without gating.  These results confirm our ability to
produce proper motions of weak pulsars that are accurate to less than
five mas yr$^{-1}$.

\section{Conclusion}

We have developed a new technique to calculate proper motions of weak
pulsars at the VLA.  By gating the VLA in one polarization, distant,
high $z$ pulsars with flux densities as low as $\sim$1 mJy can be
detected with a high signal-to-noise ratio. Previous work by
\citet{lyn82} and \citet{har93} also utilized the advantages of
in-beam reference sources and gating.  These observations were limited
by a small number of baselines (one baseline in \citet{lyn82} and two
baselines in \citet{har93}) and the positions determined using fringe
rate mapping rather than imaging could not correct for extended
structure in reference sources.  In addition, their data was degraded
by the ionosphere which is worse at 408 MHz than in our observations at
1452.4 MHz.  The 351 baselines and larger total collecting area of the
VLA enable detection of much weaker sources and improved imaging.  The
addition of wide field imaging and a detailed understanding of
systematic effects at the VLA also makes this technique more useful.
Wide-field imaging enables sources located up to $\sim20'$ from the
field center to be used as reference sources.  By imaging the
reference sources, we can exclude sources which show extended
structure.  These sources are very important in the calculation of
plate solutions as well as the pulsars' motions.  Proper motions of
the pulsars are fit simultaneously to all reference sources in the
field.

Both B0919+06 and B1237+25 agree well with previously published proper
motions confirming the accuracy of this technique.  The measurement of
the proper motion of B1937--26 shows the real success of the method.
Ungated, this pulsar has a flux density of only $\sim$1 mJy at 20 cm.
This weak flux density excluded it from all previous proper motion
projects.  By gating the VLA, proper motions of faint pulsars can now
be obtained with accuracies of a few mas yr$^{-1}$ in less than ten years. 
These accuracies
are comparable to those obtained with VLBA observations.  Although the
larger synthesized beam of the VLA reduces positional accuracies, the
VLA has the advantage of more straightforward data reduction and more
thoroughly understood systematic effects.  The presence of reference
sources in the field of view of the VLA images also makes this a
favorable technique since VLBA observations generally require
out-of-beam reference sources.  Therefore, proper motions of faint
pulsars can now be obtained at the VLA in just a few years.  Proper
motions for all 28 pulsars in this study will be reported in Paper II
by Brisken et al.

\acknowledgements{The National Radio Astronomy Observatory is a
facility of the National Science Foundation operated under cooperative
agreement by Associated Universities, Inc.  This research was funded
in part by the National Science Foundation including the REU program
at NRAO-Socorro and a graduate student fellowship.  Additional funding
was provided by Caltech and Princeton Universities.  S. E. Thorsett is
a Sloan Research Fellow.  We would like to thank Leonia Kogan for help
with the new version of UVFIX, Liese van Zee for help in acquiring the 
HI data and Ed Fomalont for advice.}

\newpage
CAPTIONS

\figcaption[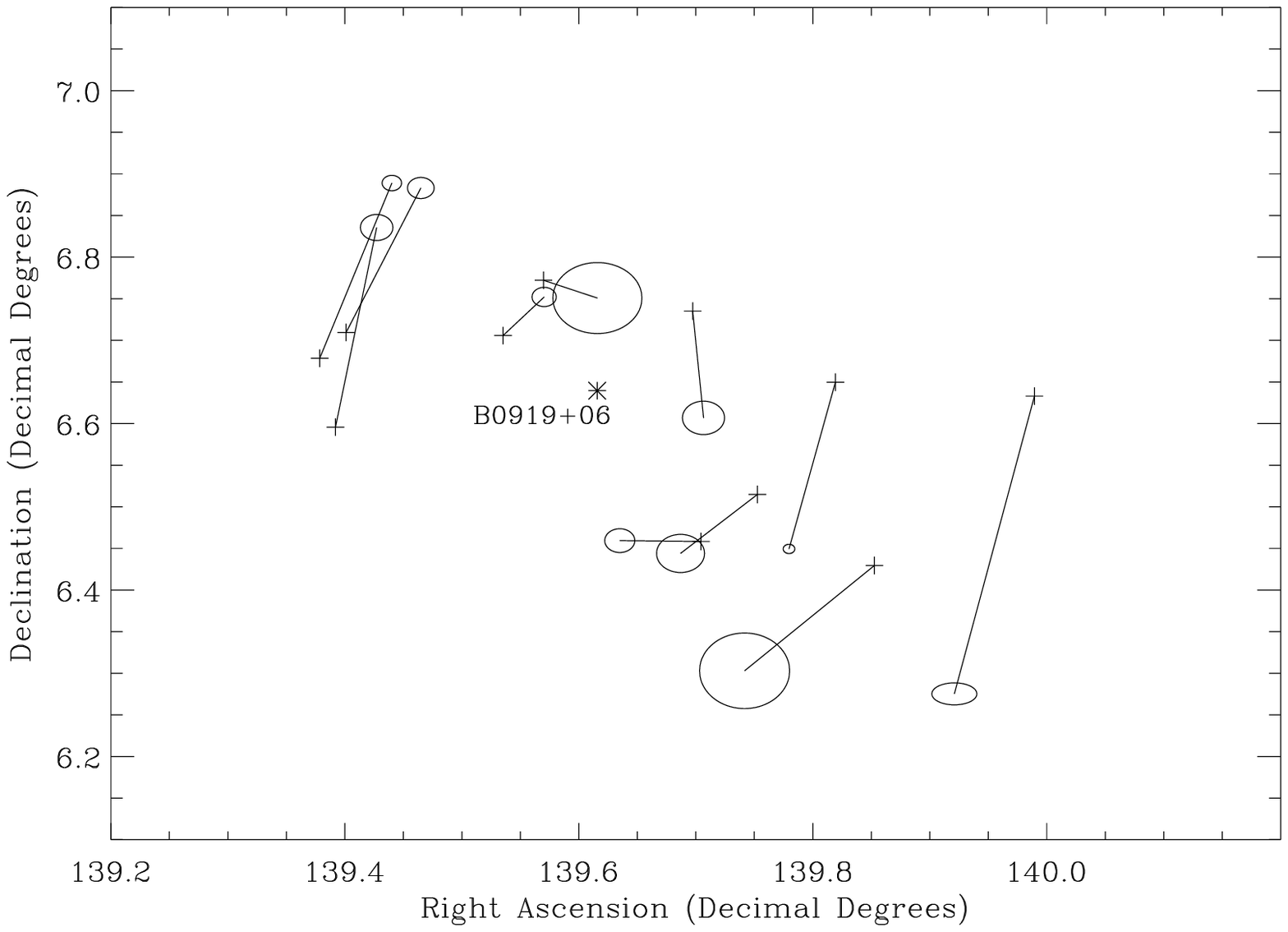]{Corrections to reference source positions as a result of the application of UVFIX.  Pulsar B0919+06 is denoted by a star while crosses represent the location of reference sources before the application of UVFIX. Ellipses represent the 1$\sigma$ error in the final position of the reference sources after UVFIX has been applied.  The correction vectors have been increased in amplitude by a factor of 5000.  The effects of annual aberration are seen in the counter-clockwise directions of the corrections about the field center. \label{uvfix:plot}}

\figcaption[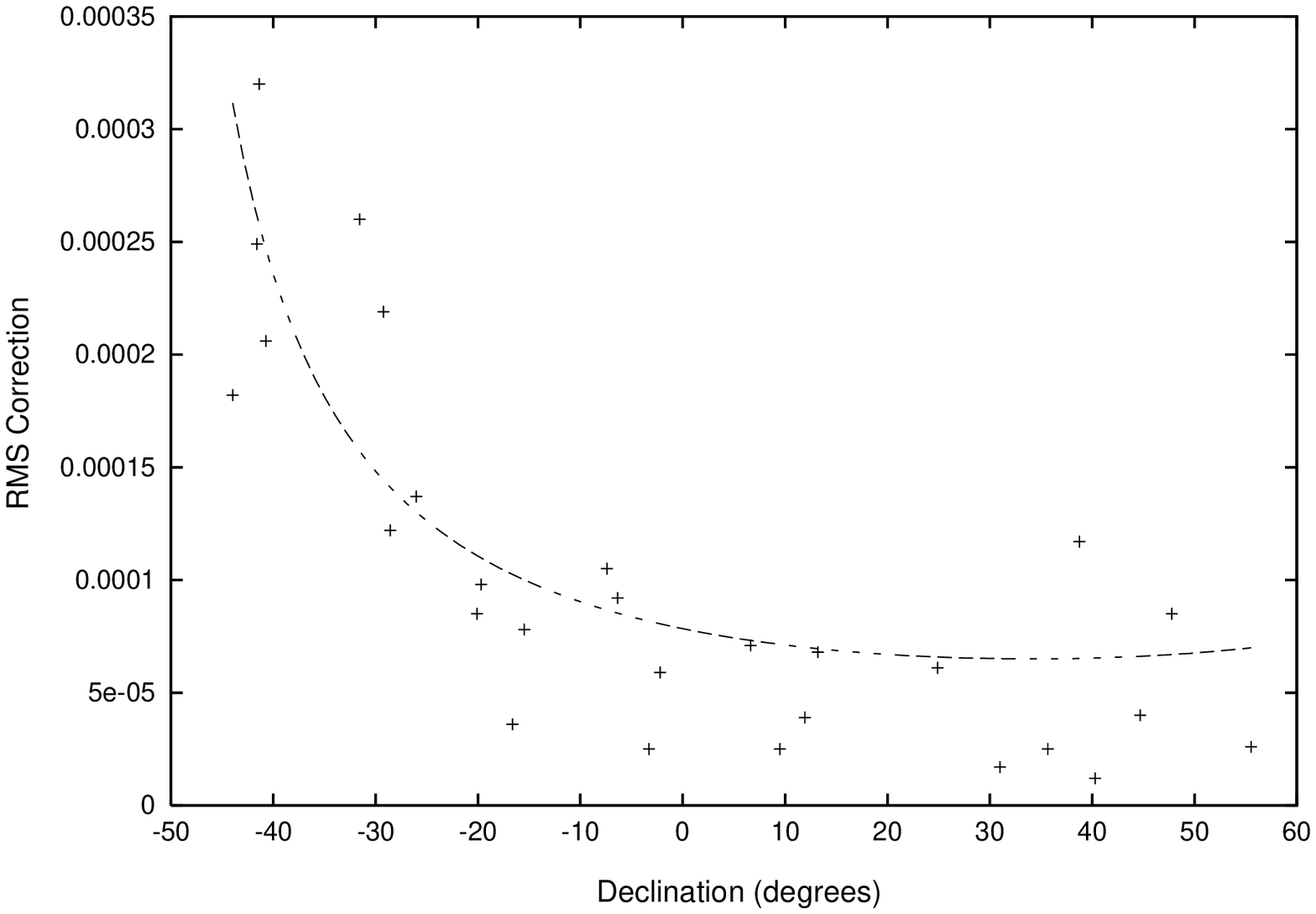] {The RMS correction required to align images
from different epochs is marked by crosses.  Since most of the
observations were made near transit the elevation is approximately $90
- \left|dec - lat \right|$, where $lat$ is the lattitude of the VLA,
about $+34^{\circ}$.  Note that the amount of
correction is approximately proportional to the secant of the
elevation (shown as the dashed line).  This implies that the cause of
the image misalignment is probably the atmosphere.  The RMS plotted here is
the fractional distortion defined in Section \ref{sec:motions}
\label{fig:elevation}}

\figcaption[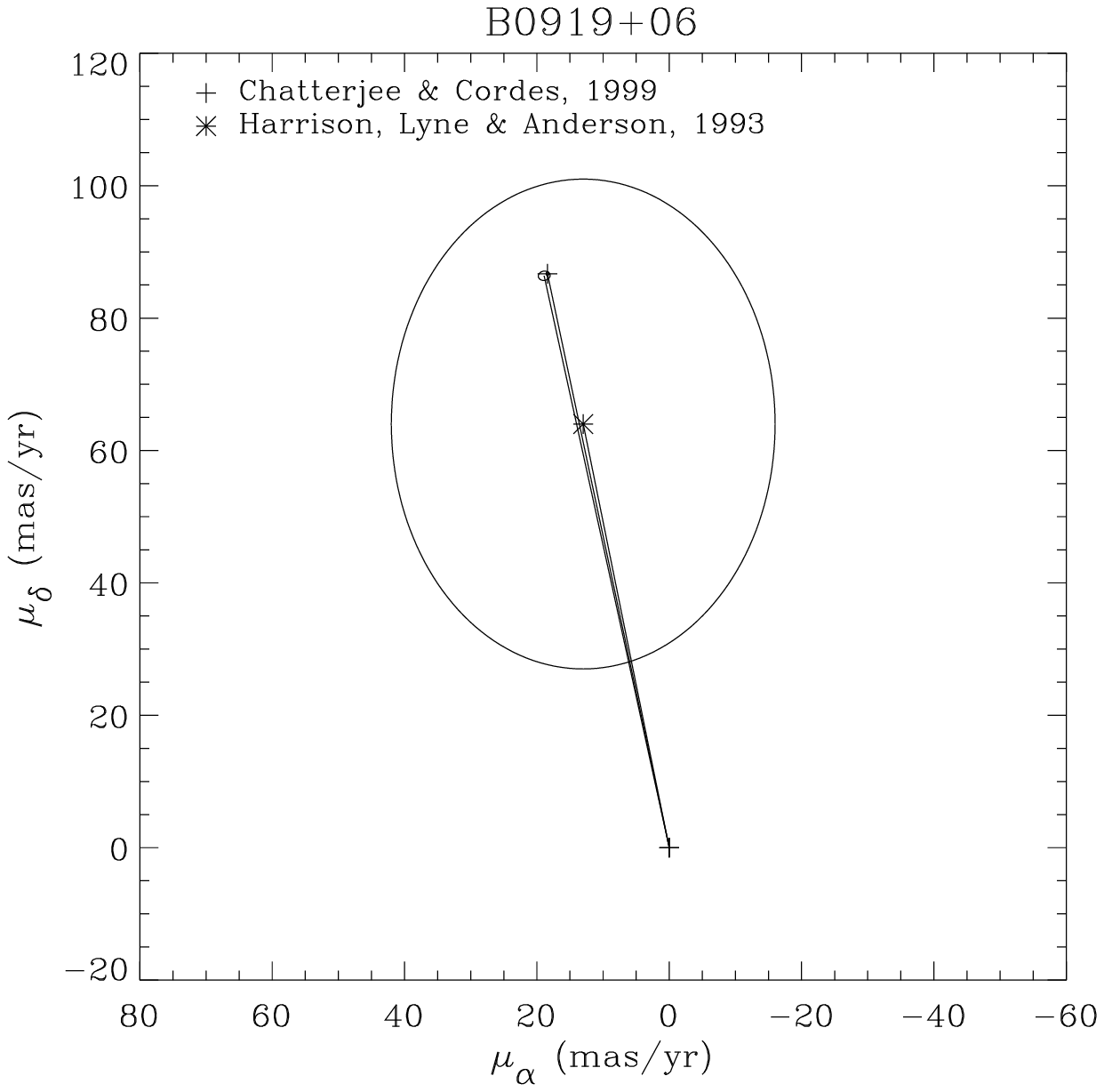]{Proper motions for B0919+06, B1237+25
and B1937--26 are shown as vectors from the origin with the 1$\sigma$
result denoted by ellipses.  For B0919+06, the \citet{chapre} result
is denoted by a cross and the \citet{har93} result is labeled by a
star.  Note that the Chatterjee et al.  error ellipse is smaller than
the size of the cross.  For B1237+25, the \citet{fom92} result is
denoted by a cross while the \citet{lyn82} result is denoted by a
star.  In all three plots, the result presented in this paper has no
symbol at the center of the ellipse. \label{pm:plot}}

\begin{deluxetable}{lrrrrrrr}
\tabletypesize{\small}
\tablewidth{0pt}
\tablecaption{Reference Sources for Proper Motions \label{refsrc}}
%\tablenum{2}
\tablehead{ \colhead{Pulsar} &\colhead{S} &\colhead{$\alpha_{2000}$} & \colhead{$\delta_{2000}$} &\colhead{$\Delta\alpha$} &\colhead{$\Delta\delta$} &\colhead{$\mu_\alpha$} &\colhead{$\mu_\delta$ }\\
\colhead{} & \colhead{mJy}& \colhead{} & \colhead{} & \colhead{arcsec} & \colhead{arcsec} & \colhead{mas yr$^{-1}$} & \colhead{mas yr$^{-1}$}}
\startdata
B0919$+$06 & 23.0 \p 0.1   &09\h23\m03\s.90 \p .01 &06\dg38\am58\as.6 \p .1     & 743 & 37  & 1 \p 4 & 1 \p 1 \\
         &   3.0 \p 0.5    &09\h22\m42\s.36 \p .01 &06\dg48\am40\as.3 \p .1 & 421 & 618  \\
         & 4.7 \p 0.1   &  09\h22\m40\s.34 \p .01 & 06\dg44\am06\as.5 \p .1 & 392 & 345   &  $-$5 \p 3  &  0 \p 3 \\  
         & 2.6 \p 0.1   &  09\h22\m12\s.08 \p .01 & 06\dg46\am19\as.9 \p .1 & $-$28 & 478   &   0 \p 6  &  2 \p 7 \\ 
         & 8.9 \p 0.1   &  09\h21\m59\s.12 \p .01 & 06\dg42\am20\as.5 \p .1 & $-$221 & 239  &   4 \p 2  & $-$1 \p 2 \\ 
         & 7.9 \p 0.1   &  09\h21\m26\s.95 \p .01 & 06\dg42\am34\as.0 \p .1 & $-$701 & 252  &  $-$1 \p 4  &  2 \p 2 \\ 
        & 10.6 \p 0.1   &  09\h21\m19\s.36 \p .01 & 06\dg40\am42\as.3 \p .1 & $-$814 & 140  &   2 \p 4  & $-$3 \p 2 \\ 
         & 6.5 \p 0.1   &  09\h21\m16\s.97 \p .01 & 06\dg35\am44\as.6 \p .1 & $-$850 & $-$157 & $-$12 \p 5  &  0 \p 2 \\ 
        & 10.0 \p 0.1   &  09\h22\m23\s.19 \p .01 & 06\dg27\am30\as.1 \p .1 & 137 & $-$652  &   1 \p 2  &  2 \p 4 \\ 
         & 4.8 \p 0.1   &  09\h22\m38\s.63 \p .01 & 06\dg30\am53\as.1 \p .1 & 367 & $-$449  &  $-$3 \p 4  & $-$1 \p 4 \\ 
         & 1.7 \p 0.2   &  09\h21\m01\s.42 \p .01 & 06\dg36\am23\as.9 \p .1 & $-$1081 & $-$118 &    \\ 
         & 3.0 \p 0.2   &  09\h21\m52\s.62 \p .01 & 06\dg15\am17\as.3 \p .1 & $-$319 & $-$1385 &    \\ 
         & 3.1 \p 0.2   &  09\h22\m57\s.06 \p .01 & 06\dg25\am46\as.1 \p .1 & 642 & $-$756  &  $-$5 \p 7  & $-$4 \p 8 \\ 
         & 2.2 \p 0.2   &  09\h23\m43\s.21 \p .01 & 06\dg26\am16\as.0 \p .1 & 1330 & $-$726 &   \\ 
         & 9.2 \p 0.1   &  09\h23\m43\s.86 \p .01 & 06\dg37\am58\as.6 \p .1 & 1339 & $-$23  &   2 \p 7  & $-$2 \p 2 \\ 
         & 1.0 \p 0.2   &  09\h23\m13\s.18 \p .01 & 06\dg50\am24\as.2 \p .1 & 881 & 721   &  \\ 
         & 2.2 \p 0.2   &  09\h23\m28\s.34 \p .01 & 06\dg58\am55\as.1 \p .1 & 1107 & 1233 &   \\ 
         & 2.6 \p 0.2   &  09\h22\m14\s.80 \p .01 & 06\dg54\am51\as.6 \p .1 & 12 & 990    &  \\ 
\\								  
B1237$+$25 & 8.7 \p 0.1 & 12\h40\m31\s.71 \p .01 &   24\dg58\am19\as.2 \p .1 & 698 & 270  &   1\p 5  & $-$2 \p 3 \\ 
        & 26.4 \p  0.2 &  12\h41\m16\s.50 \p .01 &   25\dg01\am09\as.4 \p .1 & 1305 & 440 &   &  \\ 	  
        & 9.0  \p  0.1 &  12\h40\m09\s.24 \p .01 &   25\dg06\am22\as.1 \p .1 & 391 & 753  &   1\p 3  &  4 \p 6 \\ 
        & 1.6  \p  0.1 &  12\h40\m43\s.06 \p .01 &   25\dg11\am19\as.0 \p .1 & 850 & 1049 &  & \\ 	  
        & 7.5  \p  0.1 &  12\h38\m57\s.75 \p .01 &   24\dg53\am54\as.9 \p .1 & $-$581 & 5   &   1\p 5  & $-$5 \p 2  \\ 
        & 8.2  \p  0.1 &  12\h38\m30\s.04 \p .01 &   24\dg50\am32\as.1 \p .1 & $-$958 & $-$197&  $-$2\p 7  & $-$5 \p 2  \\ 
        & 0.7  \p  0.1 &  12\h39\m07\s.26 \p .01 &   24\dg51\am05\as.2 \p .1 & $-$451 & $-$164&  & \\ 	  
        & 11.5 \p  0.1 &  12\h39\m39\s.64 \p .01 &   24\dg48\am10\as.1 \p .1 & $-$11 & $-$340 &  $-$1\p 1  &  2 \p 3  \\ 
        & 2.6  \p  0.1 &  12\h40\m58\s.92 \p .01 &   24\dg49\am27\as.8 \p .1 & 1069 & $-$262&  3 \p 10 &  6 \p 6  \\ 
        & 3.0  \p  0.1 &  12\h38\m57\s.47 \p .01 &   24\dg45\am27\as.9 \p .1 & $-$585 & $-$502&  $-$3\p 6  & $-$9 \p 5  \\ 
        & 1.8  \p  0.1 &  12\h39\m47\s.36 \p .01 &   24\dg42\am38\as.8 \p .1 & 95 & $-$671  &   5\p 5  &  9 \p 8  \\ 
        & 2.0  \p  0.1 &  12\h40\m30\s.14 \p .01 &   24\dg50\am37\as.9 \p .1 & 677 & $-$192 &  $-$4\p 7  &  6 \p 6  \\ 
\\	
B1937$-$26 & 43.0 \p  0.1 & 19\h41\m23\s.61 \p .01 &  $-$26\dg01\am15\as.6 \p .1 &  313 &   50  &  0 \p 4 &   2 \p 3  \\ 
        & 4.4   \p  0.1 & 19\h40\m48\s.26 \p .01 &  $-$25\dg58\am20\as.9 \p .1 & $-$164 &  225  & $-$3 \p 4 &   6 \p 6  \\ 
        & 2.6   \p  0.1 & 19\h40\m27\s.74 \p .01 &  $-$25\dg59\am28\as.2 \p .1 & $-$441 &  157  &  7 \p 6 &  $-$2 \p 9  \\ 
        & 3.7   \p  0.1 & 19\h39\m59\s.37 \p .01 &  $-$25\dg59\am12\as.4 \p .1 & $-$826 &  173  & $-$3 \p 9 &  $-$6 \p 8  \\ 
        & 3.2   \p  0.1 & 19\h41\m27\s.42 \p .01 &  $-$26\dg08\am05\as.4 \p .1 &  364 & $-$359  & $-$3 \p 6 &  $-$4 \p 10 \\ 
        & 4.7   \p  0.1 & 19\h42\m04\s.81 \p .01 &  $-$25\dg55\am06\as.9 \p .1 &  869 &  419  &  4 \p 9 &  $-$4 \p 8  \\
        & 2.6   \p  0.1 & 19\h42\m02\s.76 \p .01 &  $-$25\dg56\am16\as.9 \p .1 &  841 &  349  &  \\ 
        & 1.2   \p  0.1 & 19\h40\m08\s.65 \p .01 &  $-$26\dg01\am08\as.0 \p .1 & $-$698 &   56  &  \\ 
        & 18.0  \p  0.1 & 19\h40\m47\s.06 \p .01 &  $-$26\dg24\am50\as.9 \p .1 & $-$179 & $-$1365 &  \\ 
        & 2.7   \p  0.1 & 19\h42\m09\s.80 \p .01 &  $-$26\dg14\am34\as.2 \p .1 &  934 & $-$748  &  \\ 
        & 0.1   \p  0.1 & 19\h41\m43\s.25 \p .01 &  $-$26\dg05\am43\as.1 \p .1 &  577 & $-$217  &  \\ 
        & 0.1   \p  0.1 & 19\h41\m40\s.93 \p .01 &  $-$26\dg04\am51\as.1 \p .1 &  546 & $-$165  &  \\
        & 1.8   \p  0.1 & 19\h40\m20\s.86 \p .01 &  $-$26\dg10\am56\as.0 \p .1 & $-$532 & $-$530  &  6 \p 8 &  2 \p 14 \\
        & 1.3   \p  0.1 & 19\h41\m03\s.46 \p .01 &  $-$25\dg48\am33\as.6 \p .1 &   41 &  812  &  \\
\tablecomments{Sources without proper motions were not used in the determination of pulsar proper motions.}
\enddata				
\end{deluxetable}

\begin{deluxetable}{l r r l c r r c } 
\tabletypesize{\small}
\tablewidth{0pt}
\tablecaption{Proper Motions \label{pm:table}}
%\tablenum{3}
\tablehead{ \colhead{PSR} & \colhead{S} & \colhead{Gated S} &\colhead{Epochs} &  \colhead{Num. of} & \colhead{${\mu}_\alpha$} 
& \colhead{${\mu}_\delta$}& \colhead{Cov(${\mu}_\alpha$,${\mu}_\delta$)}\\
\colhead{} & \colhead{mJy} & \colhead{mJy} &\colhead{} &  \colhead{Ref.} & \colhead{mas yr$^{-1}$} & \colhead{mas yr$^{-1}$} }
\startdata
B0919$+$06 &  5.7 \p 0.1 & 120.8 \p 1.5 & 95, 98,
99\tablenotemark{\dagger} , 99 & 11&  18.8 \p 0.9 &  86.4 \p 0.7 & 0.0 \\
\\		    	              					
B1237$+$25 &  4.6 \p 0.2 & 55.0 \p 0.7 & 92, 95, 98, 99\tablenotemark{\dagger} & 9 & -104.5 \p 1.1 &  49.4 \p 1.4 & 0.1 \\
\\
B1937$-$26 &  1.0 \p 0.2 & 14.8 \p 0.6 & 94\tablenotemark{\dagger} ,
98\tablenotemark{\dagger} , 99\tablenotemark{\dagger} & 7 &  12.1 \p 2.4 & -9.9 \p 3.8 & 0.1 \\
\tablenotetext{\dagger}{Indicates that the observation was gated}
\enddata
\end{deluxetable}

\begin{figure}
\plotone{McGary.fig1.eps}
\end{figure}
\begin{figure}
\plotone{McGary.fig2.eps}
\end{figure}
\begin{figure}
\plotone{McGary.fig3a.eps}
\end{figure}
\begin{figure}
\plotone{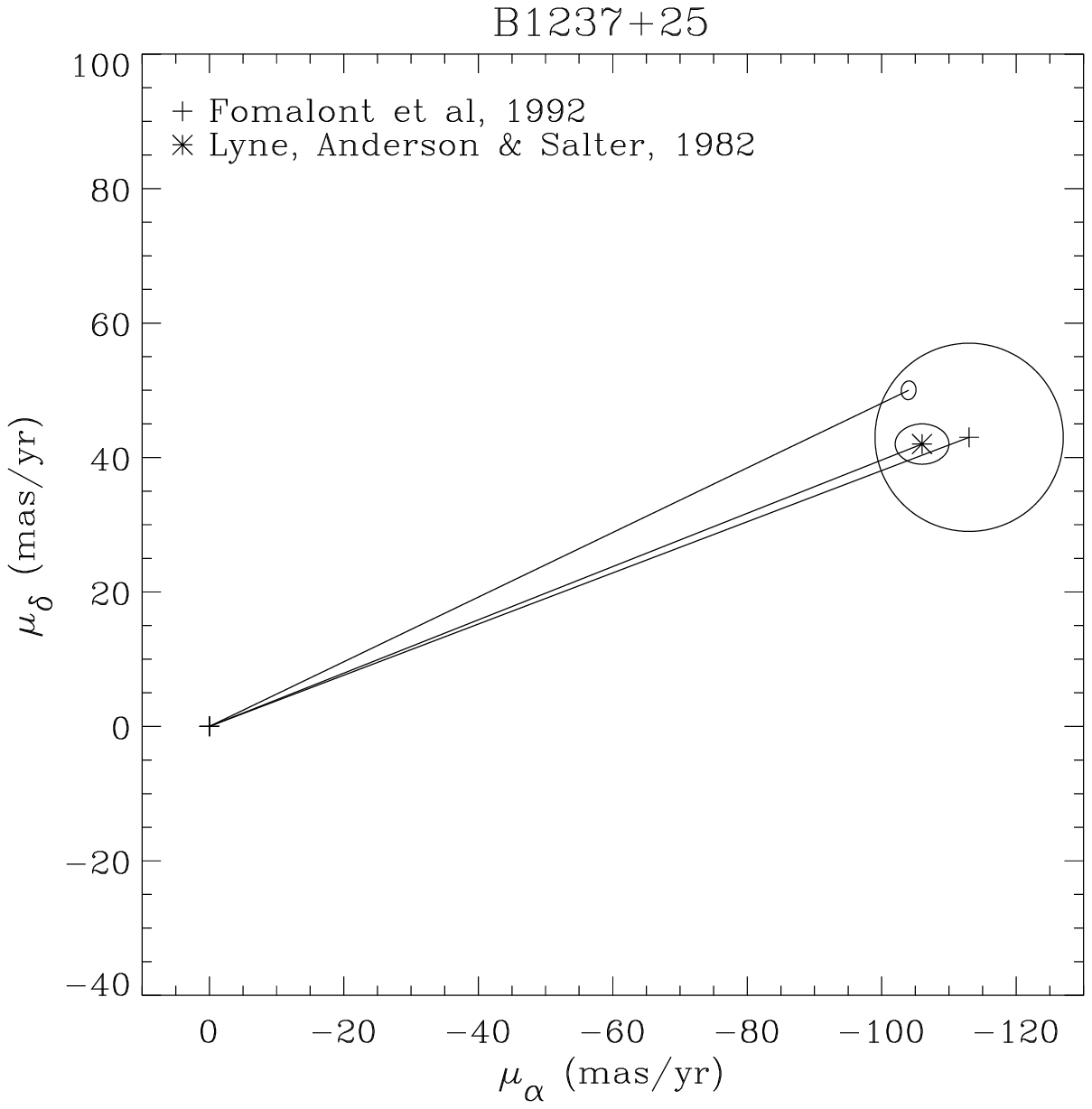}
\end{figure}
\begin{figure}
\plotone{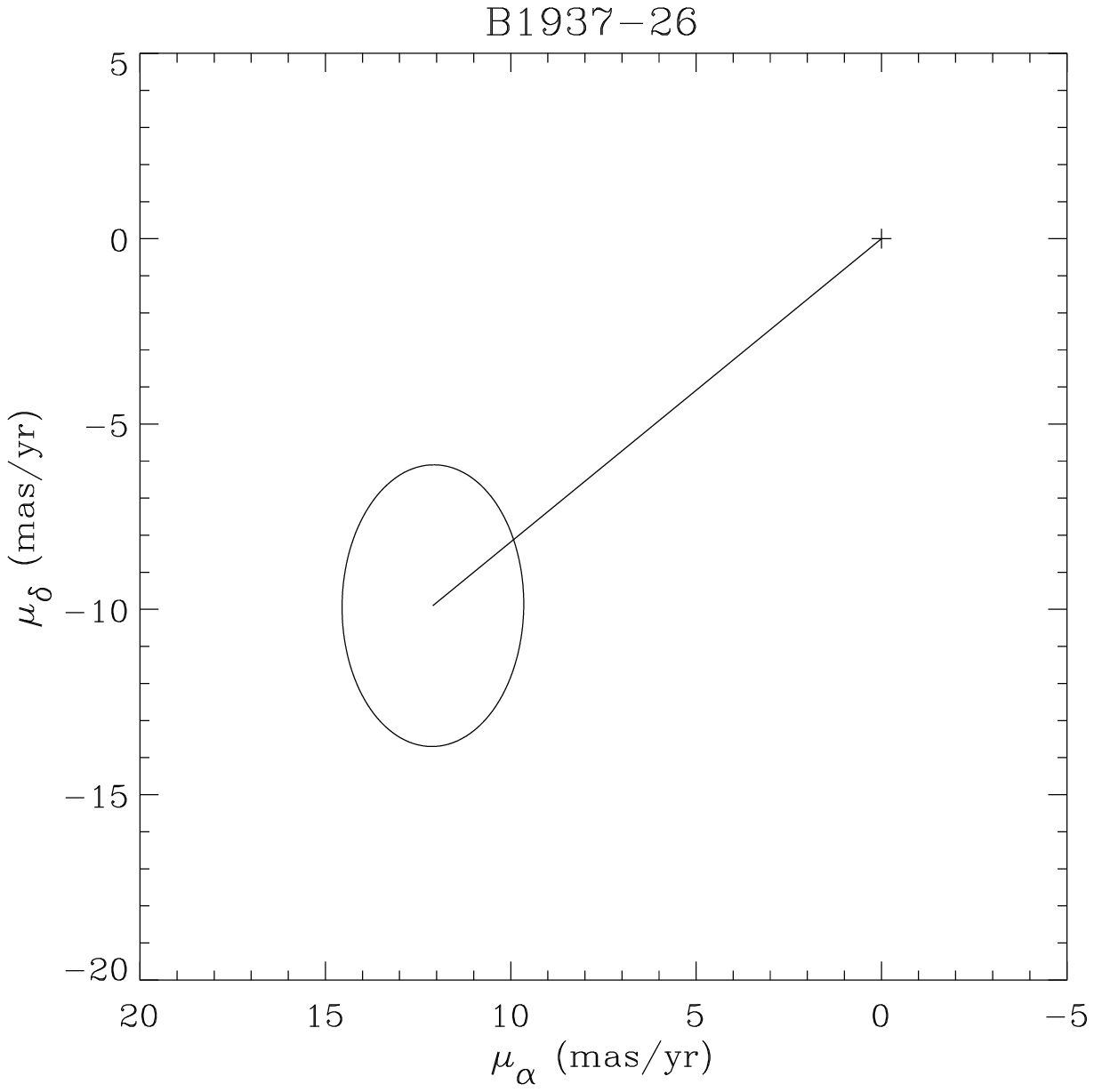}
\end{figure}

\end{document}